\documentclass[a4paper, onecolumn, notitlepage, longbibliography, floatfix, superscriptaddress, 11pt, unpublished]{quantumarticle}
\pdfoutput=1
\usepackage{geometry}
\usepackage[OT2,T1]{fontenc}
\usepackage{inputenc}
\usepackage[dvipsnames]{xcolor}
\usepackage{hyperref}
\hypersetup{
    colorlinks=true,
    linkcolor=BrickRed,
    citecolor=BrickRed,
    filecolor=VioletRed,      
    urlcolor=Plum
    }

\usepackage{graphicx}
\graphicspath{{./figures/}}
\usepackage{dcolumn}
\usepackage{bm}
\usepackage{enumerate}
\usepackage{amsmath, amssymb, amsthm}
\usepackage{faktor}
\usepackage{dsfont}
\usepackage{enumitem}

\usepackage{tabularx}
\usepackage{physics}

\theoremstyle{definition}
\newtheorem{definition}{Definition}

\theoremstyle{remark}
\newtheorem*{remark}{Remark}

\newcommand{\cP}{\mathcal{P}}
\newcommand{\cH}{\mathcal{H}}

\newcommand{\bZ}{\mathbbm{Z}}

\newcommand{\bC}{\mathbbm{C}}

\begin{document}
\title{Dynamical weight reduction of Pauli measurements} \author{Julio C. Magdalena de la Fuente}\email{jm@juliomagdalena.de}

\affiliation{Dahlem Center for Complex Quantum Systems, Freie Universit{\"a}t Berlin, Arnimallee 14, 14195 Berlin, Germany}

\maketitle
\begin{abstract}
    Many routines that one might want to run on a quantum computer can benefit from adaptive circuits, relying on mid-circuit measurements and feed-forward operations.
    Any such measurement has to be compiled into a sequence of elementary gates involving only a small number of qubits.
    In this work, we formalize \textit{dynamical weight reduction} (DWR) schemes in which a high-weight Pauli measurement is decomposed into a sequence of measurements of smaller weight at the cost of adding additional auxiliary qubits.
    We first present our main method, \textit{deforming} a ZX diagram that represents the measurement we want to compile.
    We then construct a general recipe that constructs a DWR on a given connectivity whenever the underlying connectivity graph fulfills certain necessary conditions.
    Further, we construct a family of DWR schemes using a given number of auxiliary qubits with indications that the schemes we present are optimal in terms of spacetime resource overheads needed for a DWR.
    We highlight three examples that achieve a constant time or a constant space overhead, respectively.
    Finally, we discuss different trade-offs of space and time overhead and how they might be chosen differently on different levels of abstraction within a (fault-tolerant) quantum computation.
    This work showcases the flexibility in compiling a measurement circuit in terms of lower-weight measurements using deformations of ZX diagrams and can find applications in quantum error correction, quantum simulation as well as near-term quantum computing tasks where the quality of a computation highly depends on the physical implementation of a given logical operation.
\end{abstract}


\section{Introduction}
Measurements can have a drastic effect on the physical state of a quantum system.
This ``measurement feedback'' can be used to our advantage in quantum computation, simulation and communication tasks.
For efficient state preparation~\cite{Smith2023deterministic, sahay2024finitedepthpreparationtensornetwork, stephen2024preparingmatrixproductstates, zhang2024characterizingmpspepspreparable, Bäumer2024efficient}, the read-out of error syndromes within an active quantum error-correcting or noise-reduction scheme~\cite{dennis2002topological, terhal2015review, delfosse2024clinr}, in characterization and benchmarking tasks as well as quantum communication protocols, measurements and feed-forward operations are a key ingredient to making a protocol useful and achieve a practical advantage compared to classical computations.
Additionally, recent results imply that in many instances allowing for mid-circuit measurements allow to circumvent limitations present in purely unitary implementations~\cite{Li2019measurement, Tantivasadakarn2023Hierarchy, aasen2023measurement, lootens2023lowdepthunitaryquantumcircuits}.

While most of the literature has been focussing on how to design measurements to drive systems into certain states efficiently~\cite{bravyi2022adaptive, Tantivasadakarn2023Hierarchy, Tantivasadakarn2023shortest, Iqbal2024,Bäumer2024efficient} or to perform some quantum computation~\cite{Raussendorf2001oneway, Raussendorf2007faulttolerance, Litinski2019gameofsurfacecodes} the underlying task of finding generally applicable schemes or even just heuristics to actually implement these measurements physically did not receive as much attention.
Often, the measurements that we want to implement are of high weight, meaning that they involve a number of qubits that is significantly larger than the number of qubits involved in an elementary gate on a physical device.
Hence, one has to find appropriate circuits to implement these high-weight measurements.
Depending on the device, the performance of this circuit and the overheads needed for noise-robustness, differs significantly from one circuit to the other.
That is to say, it is important to explore compilation schemes given certain constraints and elementary gate sets.

Conventional approaches to implement a generic measurement involves one extra auxiliary qubit and unitary entangling gates.
Recently, it was highlighted -- partly driven by the discovery of \textit{Floquet codes}~\cite{Hastings2021dynamically} -- that high-weight measurements can also be implemented using additional auxiliary qubits and low-weight measurements~\cite{Gidney2023pairmeasurement, GransSamuelsson2024improvedpairwise}.

We explore how high-weight Pauli measurements can be compiled into circuits that only involve one- and two-qubit measurements.
If the measurement basis is fixed, we need additional single-qubit Clifford rotations.
This type of compilation is particularly important if the elementary entangling operations on the device are measurements, such as for example in some photonic devices or topological materials~\cite{bartolucci2021fusionbasedquantumcomputation,MajoranaMilestones}.
Moreover, in many schemes for fault-tolerant quantum computation, in which a logical circuit is performed on some encoded states, two-qubit measurements are the elementary building blocks for logical entangling operations, e.g. when using lattice surgery~\cite{Horsman2012Surface,Litinski2019gameofsurfacecodes,cohen2022low-overhead,cowtan2024ssipautomatedsurgeryquantum}.

\subsection{Overview}
We present a construction on how to perform what we call \textit{dynamical weight reduction} to weight 2 (DWR$_2$).
The method we use is based on the ZX calculus, a graphical representation for operations on qubits~\cite{vandewetering2020zx}.
We introduce the concept and method of \textit{deforming ZX diagrams} in Sec.\,\ref{sec:DWR_intro}.
In Sec.\,\ref{sec:connectivity} we present the main construction of a DWR from a graph that resembles the connectivity of a device.
Given that the connectivity graph meets certain (necessary) requirements we present a DWR$_2$ scheme consisting of operations fit that connectivity graph.
We illustrate the general recipe with examples in Sec.\,\ref{sec:examples}.
In the first example, we construct a DWR$_2$ with constant space overhead, only using two auxiliary qubits, the lowest possible number for a DWR$_2$.
This comes at the cost of a time overhead (depth) that scales linearly with the weight of the measurement.
The second set of examples achieves the opposite in that it achieves DWR$_2$s with constant time overhead at the cost of a overhead of auxiliary qubits linear in the weight of the measurements.
We construct two constant-depth schemes, one of which is a reformulation of a depth-5 scheme that was also recently presented in Ref.~\cite{moflic2024constantdepthimplementationpauli} and the second one which highly improves the auxiliary qubit overhead, approximately halving it, and only increases the depth to 6.
We argue that in any DWR$_2$ a linear spacetime overhead is unavoidable so in practice one needs to choose the right trade-off.
More generally, we present an interpolating scheme in Sec.\,\ref{sec:interpolating} which achieves a low-overhead DWR$_2$ with a given number of auxiliary qubits.
In the end, we give some concluding remarks where we comment on applications and possible refinements of DWR$_2$ schemes.
We also touch on the effect of errors within a DWR scheme and sketch possible adaptations of the schemes presented that allow for more flexibility in the compilation.

\section{Dynamical weight reduction}\label{sec:DWR_intro}
Throughout this note we consider the \textit{single-qubit Pauli group} $\cP$ as a group of $2\times 2$ complex matrices generated by $i\mathds{1}$ and
\begin{align}\label{eq:Paulis}
    X=\mqty(0 & 1\\ 1& 0)\qcomma Y = \mqty(0 & -i\\ i& 0)\qq{and} Z = \mqty(1 &0\\ 0&-1).
\end{align}
$X,Y$ and $Z$ are Hermitian, square to $\mathds{1}$ and have eigenvalues $\pm 1$.
For a multi-qubit system, $\cH = (\bC^{2})^{\otimes n}$, the \textit{$n$-qubit Pauli group} $\cP_n = \cP^{\otimes n}$ is the group generated by tensor products of the matrices above.

Consider a Hermitian Pauli operator supported on $w$ qubits, $P=P_1\otimes P_2\otimes ...\otimes P_w$.
The $w$ qubits on which this operator acts can be thought of being embedded into a larger system of $n>w$ qubits.
In the following, we only consider the (sub)space of $w$ qubits on which $P$ acts on non-trivially.
The resulting operator on the whole system is obtained by tensoring with the identity operator on the rest of the system.

Measuring $P$ and with information about the measurement outcome $m\in\{0,1\}$ amounts to applying the rank-$n/2$ projector
\begin{align}
    \Pi_P(m) = \frac{1}{2}\left(\mathds{1} + (-1)^m P \right).
\end{align}
States in the image of $\Pi_P(0)$ and $\Pi_P(1)$ are unitarily equivalent.
Hence, we can without loss of generality consider the projector $\Pi_P := \Pi_P(0)$.
We will come back to fixing the sign correctly when considering adaptive Pauli corrections based on the observed outcome.
In general, the space $\Im(\Pi_P(0))$ can be mapped to $\Im(\Pi_P(1))$ by applying the Pauli unitary $Q$ that anticommutes with $P$, i.e. $PQ = -QP$.\footnote{The unitary $Q$ can be thought of as a \textit{destabilizer} of $P$. Depending on the choice of $Q$ different states in $\Im(\Pi_P(1))$ get identified with states in $\Im(\Pi_P(0))$.}

In this work, we construct decompositions of the form
\begin{align}\label{eq:dynamical-WR-def}
    \Pi_P = \Pi_{p_1}\Pi_{p_2}...\Pi_{p_k}.
\end{align}
Each $\Pi_{p_i}$ is a projector onto the $0$ outcome of $p_i$, a Pauli operator of maximal weight 2.
Importantly, $p_i$ might act on a bigger system of qubits that were not part of the $w$ qubits $P$ acts on, e.g. they form an ``environment'' of auxiliary qubits.
We refer to the qubits on which $P$ acts with a Pauli operator that is not proportional to the identity as \textit{physical qubits} and the additional qubits that we need to add in order to construct the sequence of projectors $\{\Pi_{p_i}\}_i$ as \textit{auxiliary qubits}.
At the same time, we want to point out that while this distinction makes sense for the measurement of $P$ as a whole, within the measurement, i.e. after parts of the $\Pi_{p_i}$s have been applied, there is no difference between physical and auxiliary qubits: Together, they will be non-trivially entangled.
We can view each projector $\Pi_{p_i}$ as coming from measuring $p_i$ and an appropriate (Pauli) correction to fix the right eigenspace, depending on the observed outcome.
In an architecture in which both one and two-qubit operations are native, i.e. parts of an elementary gate set, a decomposition of the form Eq\,\eqref{eq:dynamical-WR-def} corresponds to a ``compilation'' of the circuit element that measures $P$ into these low-weight, native, measurements.
For other architectures other compilations, also involving unitary entangling gates might be more useful.
But even there, decomposing high-weight measurements into lower-weight ones can be helpful, for example, when considering how errors propagate through the circuits or how syndrome information for error-correction or mitigation is gathered.

Since in general $\Pi_{p_i}$ and $\Pi_{p_j}$ don't commute the order in which the projectors are applied, i.e. the low-weight measurements are applied, is important.
Hence, we call a decomposition of the above form \textit{dynamical weight reduction} of weight 2 (DWR$_2$).

In principle, there are infinitely many such decompositions.
For example, consider a teleportation circuit composed only of measurements,
\begin{align}
    \raisebox{-0.4\height}{\includegraphics[height=2cm]{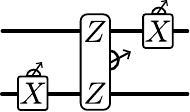}}.
\end{align}
This circuit diagram represents the following sequence of measurements on two qubits: 1. measure $X_2$, 2. measure $Z_1Z_2$ and 3. measure $X_1$.
Up to Pauli corrections that can be inferred from the measurement results, this measurement sequence ``teleports'' the quantum state on the upper qubit onto the lower qubit.
Given any DWR$_2$ of $P$ one can always add two such teleportation circuits to get a new DWR$_2$ that is longer.
One can think of more complicated decompositions that effectively differ by some ``trivial'' measurement sequence but discussing these equivalences more formally goes beyond the scope of this paper.

In the following, we want to focus on how to systematically obtain ``good'' decompositions in that they reduce certain quantities, such as the space-overhead (number of auxiliary qubits needed in the DWR$_2$) or the time-overhead (number of non-parallelizable measurements).
We will contrast these overheads in our constructions and find that any DWR$_2$ of $\Pi_P$ has a minimal spacetime overhead linear in the weight of the measurement.

Consider a circuit that implements the measurement of a weight-$w$ (Hermitian) Pauli operator $P_1\otimes P_2\otimes ... \otimes P_w$,
\begin{align}\label{eq:P-measurement-blackbox}
    \raisebox{-0.4\height}{\includegraphics[width=2.5cm]{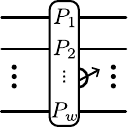}}  = \raisebox{-0.4\height}{\includegraphics[width=2.5cm]{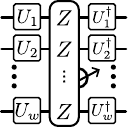}},
\end{align}
where we introduced a (Clifford) unitary $U = U_1\otimes U_2\otimes ...\otimes U_w$ that rotates the qubits into a basis where $P$ looks like a product of $Z$ operators.
For example if $P_i = X$, a possible choice is the Hadamard gate $U_i= H = (X+Z)/2$.
Since we can always find such unitaries we only have to consider a DWR of a $Z^{\otimes w}$ measurement.
This DWR can then be conjugated with $U$ to get the desired measurement circuit.

\subsection{Method: Deforming ZX diagrams}\label{sec:method}
In order to construct different DWR$_2$ we use the ZX calculus~\cite{vandewetering2020zx}.
More specifically, we start with a graphical representation of $\Pi_P$ and find different ways to ``deform'' the diagram such that it can be interpreted as a sequence of measurements leading to a DWR$_2$, cf. Eq.\,\eqref{eq:dynamical-WR-def}.
The same method was used in Refs.~\cite{Bombin2024unifying,teague2023floquetify,bauer2024lowoverhead,bauer2024xyfloquet} to derive so-called \textit{Floquet codes} from ZX diagrams representing repeated measurements of stabilizers of a topological quantum error-correcting codes~\cite{dennis2002topological} and in Refs.~\cite{Gidney2023pairmeasurement, McEwen2023relaxinghardware} to compile stabilizer read-out circuits.
Ref.~\cite{Gidney2023pairmeasurement}, in fact, performs exactly what we call a DWR$_2$.
Similarly, the ZX calculus has been very successful to formalize circuit compilation tasks.

For Pauli measurements, we only use the Clifford subcategory of the (universal) ZX calculus.
Each ZX diagram can be though of as a tensor network built from the following tensors:
A \textit{signed $\delta$-tensor},
\begin{align}
    \raisebox{-0.4\height}{\includegraphics[height=1.25cm]{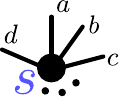}} = \begin{cases}
    (-1)^s & a=b=c=...=d\\
    0 & \text{else}
    \end{cases}\qcomma s\in\{0,1\},
\end{align}
and a \textit{signed} $\oplus$-tensor,
\begin{align}
    \raisebox{-0.4\height}{\includegraphics[height=1.25cm]{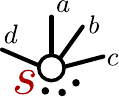}} = \begin{cases}
        1 & a+b+c+...+d = s \mod 2\\
        0 & \text{else}
        \end{cases}\qcomma s\in\{0,1\},
    \end{align}
and two-legged unitary tensors,
\begin{align}\label{eq:HandS-def}
    \raisebox{-0.4\height}{\includegraphics[height=0.4cm]{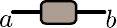}} = \frac{1}{2}(-1)^{ab} \qq{and}  \raisebox{-0.4\height}{\includegraphics[height=0.4cm]{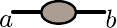}} = i^a \delta_{a,b},
\end{align}
where $\delta$ represents the \textit{Kronecker delta}, the entries of the identity matrix.
We usually omit the sign label ``$s$'' on the $\delta$- and $\oplus$-tensors if $s=0$.
One can straightforwardly check that the tensors fulfill the relations
\begin{align}
    \raisebox{-0.4\height}{\includegraphics[height=1cm]{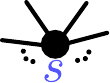}} = \raisebox{-0.4\height}{\includegraphics[height=1cm]{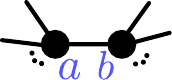}} \qq{and}
    \raisebox{-0.4\height}{\includegraphics[height=1cm]{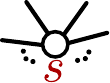}} = \raisebox{-0.4\height}{\includegraphics[height=1cm]{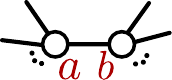}}
\end{align}
with $s=a+b\mod 2$, for any number of outgoing legs.
As a special case, we can ``split off' a tensor of the same color from a multi-legged tensor, e.g.
\begin{align}
    \raisebox{-0.4\height}{\includegraphics[width=1.5cm]{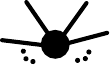}} = \raisebox{-0.4\height}{\includegraphics[width=1.5cm]{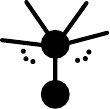}}.
\end{align}
Moreover, the tensors obey the \textit{identity rule},
\begin{align}
    \raisebox{-0.4\height}{\includegraphics[width=1.5cm]{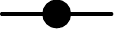}} = \raisebox{-0.4\height}{\includegraphics[width=1.5cm]{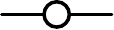}} = \raisebox{-0.3\height}{\includegraphics[width=1.5cm]{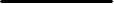}}
\end{align}

We will later identify parts of a diagram as the projector of the form $\Pi_{p_i}$.
For example, the projector $(\mathds{1} + (-1)^m ZZ)/2$ can be represented by the diagram
\begin{align}
    \raisebox{-0.4\height}{\includegraphics[width=1.75cm]{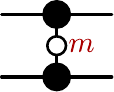}}
\end{align}
read from left to right, as we are used to in a circuit diagram.
Similarly, the projector $(\mathds{1} + (-1)^m XX)/2$ can be represented by
\begin{align}
    \raisebox{-0.4\height}{\includegraphics[width=1.75cm]{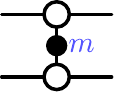}}.
\end{align}
Two-qubit measurements in other bases, such as $YY$, $YZ$, etc.~can be represented by conjugating the tensors above by unitaries defined in Eq.\,\eqref{eq:HandS-def}, representing the Hadamard and the $S = \sqrt{Z}$ gate.

Each single-qubit measurement, a projection of a single qubit into a Pauli eigenstate, is usually represented by tadpole diagrams, such as, for example, a $Z = +1$ projector is represented by
\begin{align}
    \raisebox{-0.125\height}{\includegraphics[width=1.25cm]{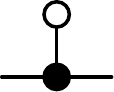}} = \raisebox{-0.4\height}{\includegraphics[width=1.25cm]{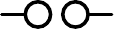}}.
\end{align}
While the diagram on the left mimics the geometry of non-destructive measurement (potentially using an additional register) the right hand side can be viewed as a destructive measurement followed by a initialization into the same state.
The equality just means that the state undergoes the same evolution in both protocols and we make no distinction between initialization in a particular single-qubit Pauli eigenstate and single-qubit measurements.
In an actual device there are, of course, differences in that the measurement can be probabilistic while a deterministic reset is not.
The key property of Pauli operators is, however, that they can be tracked efficiently through Clifford evolutions.
This makes it possible to incorporate a potentially differing measurement outcome into a Pauli correction that is performed at the end of the DWR.

For a complete introduction of the ZX calculus we refer to \cite{vandewetering2020zx}.
When analysing protocols that rely on measurements involving Pauli $Y$ operators it might be useful to add a third type of signed tensor as done in Ref.\,\cite{3colors}.

We will later use these relations to \textit{deform} the ZX diagram representing $\Pi_P$ to obtain a diagram representing a sequence of single and two-qubit measurements.
Any such deformation is composed of the following steps:
\begin{enumerate}
    \item Start with the abstract graphical (black-box) description of the linear operator that we want to implement, in our case $\Pi_P$
    \item Use splitting and joining rules such that we can...
    \item ...identify the resulting diagram as a sequence of projectors $[\Pi_{p_i}]_i$ (and possibly unitaries).
    \item Add tensors to represent other post-selections, completing the projectors to a measurement.\footnote{Here, we think of a ``Von-Neumann measurement'' being described by a set of (orthogonal) projectors that sum to the identity~\cite{nielsen2010quantum}.}
    \item Calculate Pauli correction based on the \textit{Pauli flow} of the resulting diagram
\end{enumerate}
For an introduction into Pauli flows, we refer to Ref.~\cite{3colors}.
In this context evaluating the Pauli flow representing $P$ can equivalently be thought of as a stabilizer tableau simulation of the final measurement sequence.
This tells us if the resulting circuit, given fixed measurement outcomes $\{m_i\}_i$ represents $\Pi_P(0)$ or $\Pi_P(1)$ and with that, what correction to apply, conditioned on the measurement outcomes.

Before we start with a simple example of such a deformation we want to comment that one can also think of performing DWRs of higher weight than 2 within the ZX calculus.
For one and two-qubit measurements, however, steps 3 and 4 are particularly simple.
This, and the operational simplicity of the measurement operations, are the main reasons why we focus on a full reduction to weight 1 and 2 in this work.

\subsection{Toy example: Implementing a $CX$ via measurements}
The controlled-$X$ gate is a two-qubit unitary
\begin{align}\label{eq:CNOT}
    \ketbra{0}\otimes\mathds{1} + \ketbra{1}\otimes X = (\delta_{i,i'}\delta_{j,j\oplus i})_{(ij),(i'j')},
\end{align}
where $i,i',j,j'=0,1$ label the matrix elements in the computational basis of the individual qubits.
A straight forward tensor contraction shows that it can be represented by the ZX diagram
\begin{align}
    \raisebox{-0.4\height}{\includegraphics[width=1.5cm]{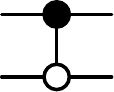}}. 
\end{align}
The upper horizontal line represents the worldline of the control qubit, i.e. the first tensor factor in Eq.\,\ref{eq:CNOT} and the lower horizonal line the worldline of the target qubit, the second tensor factor in Eq.\,\ref{eq:CNOT}.
We can use the relations of the $\delta$ and the $\oplus$ tensor to obtain an equivalent diagram
\begin{align}\label{eq:CNOT_measurements}
    \raisebox{-0.4\height}{\includegraphics[width=2.25cm]{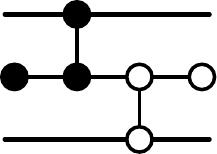}} . 
\end{align}
This diagram can be redrawn and understood as a the sequence (read from left to right):
\begin{enumerate}
    \setcounter{enumi}{-1}
    \item add one auxiliary qubit $a$
    \item project onto $X_a = +1$
    \item project onto $Z_cZ_a = +1$
    \item project onto $X_aX_t = +1$
    \item project onto $Z_a = +1$.
\end{enumerate}
Note that looking at the diagram \eqref{eq:CNOT_measurements} one can flip the middle line -- representing the worldline of the auxiliary qubit -- horizontally to end up with the same ZX diagram but interpreted as reversing the order of points 1.-4. above.
In both cases the same operator is applied in this post-selection.
As a last step of the deformation we add tensors on the vertical edges of the diagram \eqref{eq:CNOT_measurements} to represent different post-selections.
Specifically, we add two tensors and four classical bits $\{m_1,m_2,m_3,m_4\}$ that represent the measurement outcomes,
\begin{align}
    \raisebox{-0.4\height}{\includegraphics[height=2cm]{CNOT_measurements.pdf}} \quad\mapsto\quad  
   \raisebox{-0.4\height}{\includegraphics[height=2cm]{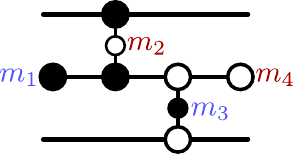}} . 
\end{align}
This diagram now represent an actual measurement circuit, i.e. a sequence of projectors $\Pi_{Z_a}(m_4)\Pi_{X_aX_t}(m_3)\Pi_{Z_cZ_a}(m_2)\Pi_{X_a}(m_1)$ which define a POVM~\cite{nielsen2010quantum}.
Note that in general the step of adding tensors to lift the post-selected ZX diagram to a diagram representing all possible post-selections is non-trivial.
For two-qubit measurements, however, the step is simple since we merely have to add a two-legged tensor on the vertical edges that connect tensors of the same type with a sign corresponding to the measurement outcome.

We can track Pauli operators from the left to the right of the diagram, graphically represented by the Pauli flows~\cite{3colors}
\begin{align}
    \raisebox{-0.4\height}{\includegraphics[width=3.2cm]{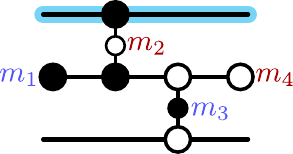}} , \raisebox{-0.4\height}{\includegraphics[width=3.2cm]{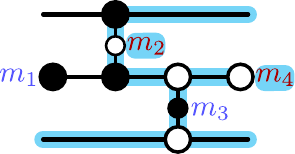}} , \raisebox{-0.4\height}{\includegraphics[width=3.2cm]{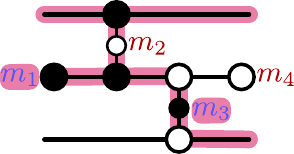}}, \raisebox{-0.4\height}{\includegraphics[width=3.2cm]{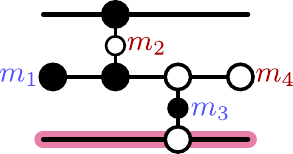}}
\end{align}
to obtain the action of this circuit on the Pauli group on the input qubits.
We find that the above diagram implements the map
\begin{subequations}
    \begin{alignat}{2}
        Z_c &\mapsto Z_c,
        &&X_c \mapsto (-1)^{m_1+m_3}X_cX_t, \\
        Z_t &\mapsto (-1)^{m_2+m_4}Z_cZ_t,\quad 
        &&X_t \mapsto X_t.
    \end{alignat}        
\end{subequations}
This differs from the action of a $CX$ gate by a Pauli operator, $Q(\vb{m}) = X_t^{m_2+m_4}Z_c^{m_1+m_3} $, where $\vb{m}=(m_1,m_2,m_3,m_4)$ represents the array of measurement outcomes.
This correction can now be applied directly, or, within a bigger circuit, tracked offline and applied ``just in time'' before the next Non-Clifford operation in order to achieve the correct (unitary) action of the measurement circuit.

\section{DWRs from connectivity graphs}\label{sec:connectivity}
In this section we present a construction of a DWR$_2$ for a weight-$w$ measurement that fit a given connectivity. Note that we only consider $w>2$ since this is the first non-trivial case in which two-body measurements need to be compiled to implement the measurement.
Besides that, there are no restrictions on the weight $w$.

The construction starts with the ZX diagram,
\begin{align}\label{eq:P-measurement-blackboxZX}
    \Pi_{Z^{\otimes w}}(m) = \raisebox{-0.4\height}{\includegraphics[width=2cm]{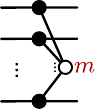}}\quad ,
\end{align}
which can be seen analogous to the circuit diagram \eqref{eq:P-measurement-blackbox}. The black dots represent the $w-3$ additional wires that are not explicitly show.

A na\"ive read-out circuit, where all entangling operations are unitaries can be represented by the following equivalent ZX diagram,
\begin{align}
    \raisebox{-0.3\height}{\includegraphics[width=3cm]{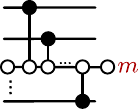}} ,
\end{align}
that is a straightforward deformation of the black-box diagram obtained from splitting the $\oplus$-tensor, cf. Sec.\,\ref{sec:method}.

In this section we present a sequence that realizes a deformed ZX diagram to the one above in which all two-qubit operations fit a given connectivity.
We formalize the connectivity of a device in terms of a \textit{connectivity graph}.
To make the construction tractable, we treat every qubit on the device on equal footing and assume that we can perform arbitrary single-qubit Pauli measurements on each qubit and arbitrary two-qubit Pauli measurements between certain pairs of qubits.
\begin{definition}[Connectivity graph]
    The \emph{connectivity graph} of a device is a graph whose vertices are in one-to-one correspondence with the qubits on the device.
    Two vertices $a$ and $b$ are connected by an edge $(a,b)$ if two-qubit Pauli measurements can be performed between the qubits associated to $a$ and $b$.
\end{definition}
We give a construction of a DWR$_2$ on a subgraph of the connectivity graph $G$ that fulfills the following conditions:
\begin{itemize}
    \item has at least $w+2$ vertices,
    \item admits a spanning tree $G_s$ with $w$ leaves.
\end{itemize}
The qubits assigned to the leaves of $G_s$ take the role of the physical qubits and the qubits assigned to the remaining vertices of $G$ act as auxiliary qubits for the measurement.
We denote its set of leaves of $G_s$ with $L$, $P$ the set of it's parents. The remaining vertices of $G$ are denoted with $I$.
Note that the full vertex set of $G$ is the disjoint union of $L,P$ and $I$.
Sometime, we will refer to the leaves connected to a given element $p\in P$ as \textit{children} of $p$.

\begin{figure}
    \centering
    \includegraphics[width=0.9\linewidth]{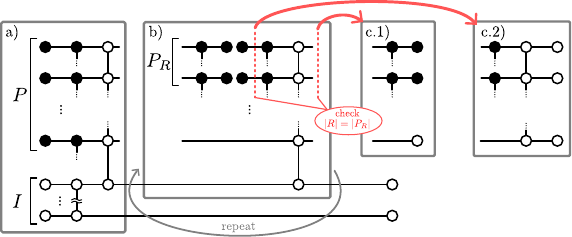}
    \caption{Illustration of the measurement sequence obtained following the general construction.
    We show parts of the resulting ZX diagram on the auxiliary qubits.
    $ZZ$ measurements, represented by trivalent $\delta$-tensors, are between  qubits in $P$ and physical qubits, not shown in the diagram. Each $XX$ measurement, indicated by $\oplus$-tensors connected by vertical edges, is amongst the auxiliary qubits.
    a)~In the first three steps all qubits in $P$ participate in one $ZZ$ measurement with a physical qubit and get entangled with all other auxiliary qubits $P\cup I$ via a sequence of $XX$ measurements.
    b)~After step 3., the construction enters a repeating process that ends once the remaining physical qubits $R$ can be assigned 1-1 to their parents $P_R$.
    Note that before each time $\abs{R}=\abs{P_R}$ is checked the sets $R$ and $P_R$ get updated so they get smaller through repeating steps 2.-5.
    c)~Depending on the steps before exiting the repeating loop, the resulting DWR ends differently, cf. c.1) and c.2).}
    \label{fig:general_construction}
\end{figure}

We can implement a $Z^{\otimes w}$ measurement on the qubits $L$ by the following measurement sequence:
First, for all $i\in I$, measure $Z_i$ followed by $X_iX_j$ for every pair $i,j\in I$ connected via an edge in $G$.
Then, for each $p\in P$,
\begin{enumerate}
    \item measure $X_p$,
    \item pick one child of $p$, $c_p\in L$ and measure $Z_pZ_{c_p}$,
    \item measure $X_pX_i$ for every neighbor $i\in P\cup I$, omitting repeated measurements of the same pair of $X$ operators.
\end{enumerate}
Let $R$ be the set of leaves that were not involved in any measurement yet.
Consider the set of parents of elements in $R$, $P_R$.\footnote{Note that in general $\abs{R}\geq\abs{P_R}$. Equality holds if every $p\in P_R$ has a unique child in $R$.}
If $\abs{P_R} = \abs{R}$, continue with step 6.
If $P_R=P\cup I$, pick an element $m\in P_R$ with the least number of children and remove $m$ from $P_R$.\footnote{The choice of which element to remove from $P_r$ is arbitrary. The choice we make here is useful to minimize the full overhead of the resulting DWR scheme.
Moreover, note that we still consider $m$ to be an element of $P$.}
For each $p\in P_R$,
\begin{enumerate}[resume]
    \item pick a child of $p$, $c_p\in R$, and measure $Z_pZ_{c_p}$,
    \item measure $X_p$.
\end{enumerate}
Update $R$ to the set of leaves that have not participated in any measurement yet and $P_R$ to the set of their parents.
As long as $\abs{P_R}< \abs{R}$, repeat starting from step 2, replacing $P$ with $P_R$ if $P_R\neq P\cup I$. In the case of $P_R= P\cup I$, remove a $m\in P_R$ with the least number of children before repeating.
\begin{enumerate}[resume]
    \item For each untouched leaf $l\in R$, pick a parent $p\in P_R$ and measure $Z_lZ_p$.
    \begin{itemize}
        \item If the measurement performed on $p\in P_R$ before the $Z_lZ_p$ measurement was a single qubit $X_p$ measurement (step 5) measure $X_pX_i$ for every neighbor $i\in P\cup I$.
    \end{itemize}
    \item For each $i\in P\cup I$, if the previous measurement it participated in was in the $X$ basis, measure $Z_i$, if it was in the $Z$ basis, measure $X_i$.
\end{enumerate}

In Fig.\,\ref{fig:general_construction} we illustrate the procedure by sketching the ZX diagram of the resulting measurement sequence on the auxiliary qubits.
\begin{remark}
    Note that for any $G$ fulfilling the above conditions, we can replace it with its spanning tree $G_s$ in the construction above and still obtain a valid DWR$_2$ for the same measurement.
    This reduces the spacetime overhead of the DWR scheme by reducing the number of redundant measurements coming from loops in $G$.
    For the purpose of error-detection it might still be useful in some cases, to involve more measurements than optimal into the DWR.
\end{remark}

\begin{figure}
    \centering
    \includegraphics[width=\linewidth]{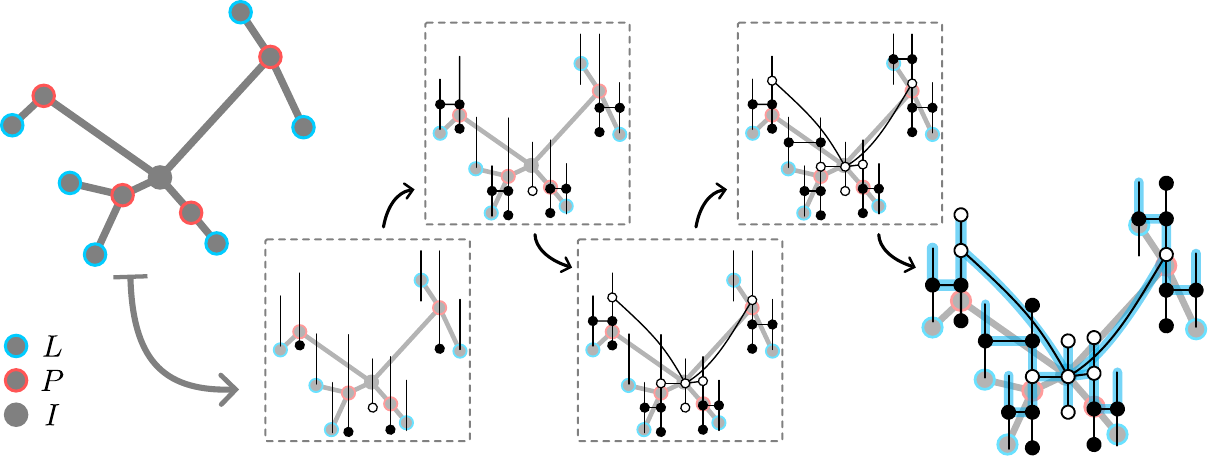}
    \caption{We provide a construction of a DWR$_2$ on a given connectivity.
    In the figure we show an exemplary construction of a DWR$_2$ for a weight-6 measurement starting with a subgraph of the connectivity graph.
    The physical qubits, associated to the leaves of the graph $L$, on which the full measurement in performed are marked in blue, their parents $P$ in red.
    Additionally, there is a single internal node $I$ not connected to any physical qubit directly.
    The graph shown is a generic graph that fulfills the necessary conditions for our construction.
    From left to right we show how the associated ZX diagram is built up during the stages of the DWR sequence.
    The right-most diagram represents the projector implemented by the DWR post-selecting on the 0 outcomes together with the Pauli flow representing the stabilizer $Z^{\otimes 6}$ on the physical qubits initialized by the DWR$_2$.
    }
    \label{fig:DWR-connectivity-graph}
\end{figure}

The depth of the resulting DWR highly depends on the details of $G$.
As a heuristic we can deduce is that if $G$ admits a ``hub'', an internal node through which many of the shortest paths in $G$ run through, this qubit has to participate in many measurements and thereby increasing the depth.
For the explicit constructions in Secs.\,\ref{sec:constant_space} and \ref{sec:constant_time} we give expressions for the depth in terms of parameters of the graphs used there.
We leave a more rigorous treatment of how graph-theoretic quantities affect the depth in the more general setting to future work.

The measurement circuit obtained from this recipe involves many measurements and with that we obtain many bits $m_i=0,1$, each of which (individually) is random, if no prior assumptions on the state of the qubits are made.
In fact, the full circuit measures $Z^{\otimes w}$ and the measurement outcome is obtained by summing up the $ZZ$ and $Z$ measurement outcomes modulo 2.

In Fig.\,\ref{fig:DWR-connectivity-graph}, we show the construction and the associated ZX diagram for an exemplary graph.
From the ZX diagram one can directly infer that the measurement sequence obtained implements the correct measurement and the Pauli flow on that diagram also gives direct way to see which measurement outcomes contribute to the sign of the measured operator.

\subsection{Necessary connectivity conditions for DWR$_2$}
For any implementation of a measurement of $w$ qubits, they have to be in a single connected component of the connectivity graph to build up the entanglement needed.
Moreover, a DWR$_2$ of a measurement of weight $w>2$ requires at least two additional qubits.
A simple way to see this is via the ZX diagram from Eq.\,\eqref{eq:P-measurement-blackboxZX} representing the measurement.\footnote{Equivalently, we could prove our statement using the stabilizer formalism and resorting to the \textit{instantaneous stabilizers} of the system throughout any DWR$_2$ scheme. To keep it simple, we only present the graphical argument here.}
The node carrying the measurement outcome is connected to exactly $w$ $\delta$-tensors (of degree 3) placed on the worldlines of the physical qubits.
Assuming that the physical qubits are fixed there are some non-trivial conditions on the diagram at step 3 (cf. Sec.\,\ref{sec:method}) of the deformation that leads to a DWR$_2$ scheme.
At this stage, each 2-qubit measurement is represented by two tensors of the same type ($\delta$ or $\oplus$) connected via an edge that does not lie on any qubit worldline.
Moreover, each edge between $\delta$- and $\oplus$-tensors has to be on a worldline of an auxiliary qubit.
If not, it would represent a $CX$ gate which we explicitly want to avoid.
From this we can directly deduce that one auxiliary qubit is not sufficient for a DWR$_2$:
Let the physical qubits of the measurement be labelled by $1,...,w$ and label the $\delta$-tensors on the respective worldlines before the deformation procedure by the same number.
Take qubit 1. After applying ZX rewrites (step 3. in the deformation step, cf. Sec.\,\ref{sec:method}) there has to be at least one edge from the a $\delta$-tensor (obtained from splitting tensor 1) to a $\oplus$-tensor.
As an edge between tensors of different types it has to correspond to a short section of the worldline of an auxiliary qubit.
As a worldline, it has to continue after the $\oplus$-tensor.
For $w>2$ there are additional edges emanating from that tensor, also after applying rewrite rules.
There are two possibilities: On the one hand, it can connect to another $\delta$-tensor, say number 2, and thereby represent a $CX$ gate between auxiliary and phyiscal qubit. On the other hand, it can connect to another $\oplus$-tensor, and correspond to a (post-selected) $XX$ measurement.
This, however, can only be with another auxiliary qubit since the only tensors on the worldlines of the physical qubits are $\delta$-tensors.
This shows in order to have all entangling operations being a measurements, there are at least two auxiliary qubits necessary to implement a measurement of weight $w>2$.
Together with the construction we present in Sec.\,\ref{sec:constant_space} we show that two auxiliary qubits are sufficient, leading to a minimal requirement of $w+2$ qubits on the device to run a DWR$_2$ scheme with native 1- and 2-qubit measurements.


Other constraints on the connectivity graph can be lifted if we allow for additional SWAP gates.
It would allow for a weight-$w$ measurement on any connected component with at least $w+2$ qubits without the necessity of $w$ leaves.
This, however, comes with a time overhead since the qubit states have to be ``moved'' onto the right physical qubits at each stage of the measurement circuit, effectively simulating a connectivity in which the physical qubits of the measurement are the leaves of a subtree.
For example, one can simulate any connectivity with additional SWAPs using only a one-dimensional connectivity.

When only allowing for measurements as two-qubit operations we find that the non-trivial constraint on the connected component involving at least $w+2$ qubits is that it has to have a spanning tree with $w$ leaves, the case we discussed above.
Whenever it has less leaves at least one of the qubits in the interior of the graph would need to be physical qubits of the measurement.
At the same time this means that this internal qubit has to participate in more than a single measurement to spread the entanglement, respectively grow the Pauli stabilizer within the DWR sequence, over the connected component.
We only want the physical qubits to participate in a single measurement in order to only perform the weight-$w$ measurement on the physical qubits.
The fact that there exists no spanning tree with $w$ leaves means that the internal qubits (of some spanning tree of the connected component) have to participate in at least two measurements, or SWAPs have to be used.

\section{Trading off space and time overheads}\label{sec:examples}
In this section we present a family of DWR$_2$s that realize different trade-offs between space and time overhead.
We first present schemes realizing two extreme cases in which constant space or constant time overheads are achieved. The first example implements a measurement of $P$ with only two additional auxiliary qubits, at the cost of circuit depth increasing linearly in the weight of $P$.
The second set of DWR$_2$s can be thought of achieving the opposite:
They realize a constant time overhead in that they implement the measurement of weight-$w$ $P$ in just 5 or 6 steps.
That is to say, they comes with a space overhead of $w$ or $\lceil w/2\rceil$ auxiliary qubits.
We note that depth-5 DWR$_2$ that achieves a constant time overhead was also presented in Ref.~\cite{moflic2024constantdepthimplementationpauli}.
We improve on that in our second constant-depth scheme where we achieve a constant depth of 6 only using around half the auxiliary qubits.
Using the recipe from Sec.\,\ref{sec:connectivity} we present a family of schemes that interpolate between the two extreme cases.
The family is labelled by the number of auxiliary qubits used $2\leq a\leq w$ and reproduces the schemes we present before for $a=2,w/2$ and $w$.
All DWRs realize a spacetime overhead that scales linearly with $w$. We end the section arguing that this overhead is unavoidable indicating that the schemes we present are (close-to) optimal with regards to resources needed for the DWRs.
Moreover, we comment on how the chosen trade-offs influence the potential parallelizability of sequential measurements involving the same physical qubits.

\subsection{Constant space overhead}\label{sec:constant_space}
In this section, we present a DWR$_2$ scheme using only $2$ auxiliary qubits.
It is obtained from the general recipe presented in Sec.\,\ref{sec:connectivity} using the graph
\begin{align}
    \raisebox{-0.4\height}{\includegraphics[height=3.5cm]{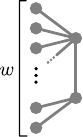}}.
\end{align}
The graph is aligned such that the $w$ physical qubits are on the left and the 2 auxiliary qubits on the right.
The bottom auxiliary qubit is only connected to two physical qubits and the remaining physical qubits are all connected to the upper auxiliary qubit.\footnote{One could also consider a graph where the connections from the auxiliary qubits are more evenly distributed amongst the auxiliary qubits which might seem more natural.
This leads to a very similar DWR, where the roles of the auxiliary qubits change throughout the circuit. Here, we choose the more asymmetric graph from which we define the DWR for a simpler presentation of the resulting circuit.}
For now, assume $w$ is even and bigger than 4.
Following the recipe in Sec.\,\ref{sec:connectivity} we obtain a measurement sequence with an associated ZX diagram
\begin{align}\label{eq:constant-space}
    \raisebox{-0.4\height}{\includegraphics[height=3.75cm]{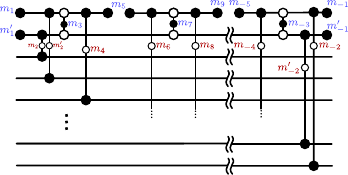}}.
\end{align}
In App.\,\ref{app:constant-depth-splitting}, we show how this sequence can be obtained from deforming the ZX diagram resembling the $Z^{\otimes w}$ measurement.
In words, labelling the physical qubits from 1 to $w$ and the auxiliary qubits by $a$ and $b$, the diagram corresponds to the measurement sequence
\begin{enumerate}
    \item Measure $\{X_a,X_b\}$, get outcomes $\{m_1,m_1'\}$.
    \item Measure $\{Z_aZ_1, Z_bZ_2\}$, get outcomes $\{m_2,m_2'\}$.
    \item Measure $X_aX_b$, get outcome $m_3$.
    \item Measure $Z_aZ_3$, get outcome $m_4$.
    \item Measure $X_a$, get outcome $m_5$.
\end{enumerate}
Let $r$ be the number of physical qubits that where not involved in a measurement yet.
As long as $r>3$, repeat
\begin{enumerate}[resume]
    \item Measure $Z_aZ_{w-r}$
    \item Measure $X_aX_b$
    \item Measure $Z_aZ_{w-r+1}$
    \item Measure $X_a$
\end{enumerate}
When $r= 3$ do the final round of measurements on the physical qubits:
\begin{enumerate}[resume]
    \item Measure $Z_aZ_{w-2}$, get outcome $m_{-4}$.
    \item Measure $X_aX_b$, get outcome $m_{-3}$.
    \item Measure $\{Z_aZ_{w-1}, Z_bZ_w\}$, get outcomes $\{m_{-2}, m_{-2}'\}$
    \item Measure $\{X_a, X_b\}$, get outcomes $\{m_{-1}, m_{-1}'\}$.
\end{enumerate}
Carefully counting the number of repetitions of steps 6.-9. we obtain a depth of $5+4\lfloor\frac{w-3}{2} \rfloor=\Theta(w)$.

Following the same procedure for $w$ being odd we find that the last timesteps look different.
Specifically, the end of the measurement circuit will not terminate both auxiliary qubits with a $\delta$-tensor but only one of the two.
The final diagram in that case will look the same but with the end taking the form (omitting the measurement labels)
\begin{align}
    \raisebox{-0.4\height}{\includegraphics[height=1.75cm]{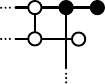}},
\end{align}
i.e. after the last $XX$ measurement only one of the two auxiliary qubits participates in a $ZZ$ measurement with a physical qubit while the other one gets read out in the $Z$ basis.

While Ref.~\cite{Gidney2023pairmeasurement} constructed a DWR$_2$ sequence for weight-4 operators (the stabilizers of the surface code) with two auxiliary qubits this is -- to the best of our knowledge -- the first construction that achieves this for a generic Pauli measurement.

By construction we know that if all measurement outcomes are 0, the circuit implements the projection $\Pi_{Z^{\otimes w}}$.
For any other post-selection the circuit projects onto some definite eigenspace of $Z^{\otimes w}$, i.e. implements $\Pi_{Z^{\otimes w}}(m)$ for some $m\in\{0,1\}$.
We can track Pauli operators through the circuit to obtain $m$ as a function of the measurement outcomes observed.
For example, we can do this by calculating the Pauli flow of the diagram,
\begin{align}
    \raisebox{-0.4\height}{\includegraphics[height=4cm]{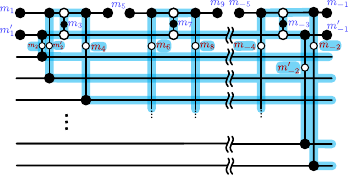} },
\end{align}
and adding up all highlighted measurement outcomes modulo 2 and find that $m$ is given as the sum of all measurements related to $Z$ measurements, i.e. in steps 2, 4, 6, 8 (and their repeated versions), 10 and 12 in the sequence described below \eqref{eq:constant-space}.
Note that the Pauli flows can be found efficiently for any ZX diagram.
Equivalently, one can think of the diagram sequentially and obtain $m$ by a standard stabilizer tableau simulation~\cite{aaronson2004improved,Gidney2021stimfaststabilizer}.

\subsection{Constant time overhead}\label{sec:constant_time}
In the previous DWR scheme we have seen that while having a constant space overhead of only two qubits the time overhead scales linearly with the weight $w$ of $P$.
We present two DWR$_2$ schemes that achieve a constant time overhead.
The lowest weight scheme we obtain has depth 5 and uses $w$ auxiliary qubits.
A second example slightly increases the depth to 6 but halves the number of auxiliary qubits needed to $\lceil w/2\rceil$.

\subsubsection{Depth-5 scheme}
The first scheme we present here is derived using the recipe from Sec.\,\ref{sec:connectivity} on the graph
\begin{align}
    \raisebox{-0.4\height}{\includegraphics[height=4cm]{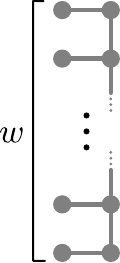}}.
\end{align}
It is built by pairing each auxiliary qubit up with a physical qubit (horizontal edges above) and then connecting the auxiliary qubits among each other along a line (vertical edges above).\footnote{The same depth is achieved using two less auxiliary qubits with the graph \raisebox{-0.4\height}{\includegraphics[height=16pt]{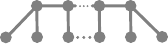}}. We will find a significantly better saving in auxiliary qubits in the next example.}
Following, the recipe from Sec.\,\ref{sec:connectivity} we obtain the DWR$_2$ represented by the ZX diagram
\begin{align}\label{eq:constant-depth}
    \raisebox{-0.4\height}{\includegraphics[height=5cm]{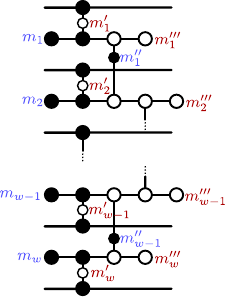}}.
\end{align}
In App\,\ref{app:constant-depth-splitting} we show how the above diagram can be obtained by deforming the ZX diagram representing the $Z^{\otimes w}$ measurement.
The above diagram represents the following measurement sequence of depth 5:
\begin{enumerate}
    \item Take $w$ auxiliary qubits. For each auxiliary qubit $a$, measure $X_a$ and get outcomes $\{m_i\}_{i=1,...,w}$.\footnote{Equivalently, one can initialize each auxiliary qubit in the $\ket{+}$-state and set $m_i=0\;\forall i$.}
    \item Pair each physical qubit $i$ with unique auxiliary qubit $a_i$ and measure $Z_iZ_{a_i}$, get outcomes $\{m'_{j}\}_{j=1,...,w}$.
    \item Measure $X_{a_i}X_{a_{i+1}}$, $i=1,...,w-1$ between neighboring auxiliary qubits, get outcomes $\{m''_k\}_{k=1,...,w-1}$. Note that each auxiliary qubit, except two, participates in two of these measurements.
    \item Measure $Z_a$ on each auxiliary qubit, get outcomes $\{m'''_{l}\}_{l=1,...,w}$.
\end{enumerate}
By looking at the diagram \eqref{eq:constant-depth} representing the protocol above we find that one could equivalently mirror it along a vertical line and still implement the same measurement (up to Pauli corrections).
The mirrored image would correspond to an equivalent protocol where one initializes in $Z$ eigenstates, entangles the auxiliary qubits by $XX$ measurements and then entangles them with the phyiscal qubits using $ZZ$ measurements.
The final read-out is then done in the $X$ basis.
This can also be interpreted as preparing the auxiliary qubits in an entangled (GHZ-like) state and performing transversal entangling measurements with the physical qubits before measuring out the auxiliary qubits again.
In principle, one has the choice of first entangling auxiliary qubits with each other and then with the physical qubits for each subset of auxiliary qubits.
This could be helpful when considering repeated measurements since the final read-out can be seen as an initialization in the same basis if the measurement can be done non-destructively.
In that way, one could directly reuse the just-measured auxiliary qubit in an entangling operation for the next measurement.

Although the depth of the above protocol is constant (5), the number of auxiliary qubits is exactly $w$ giving rise to spacetime overhead linear in $w$.

By construction the measurement sequence presented above applies $\Pi_{Z^{\otimes w}}(0)$ if all measurement outcomes are 0.
For the other cases, we calculate the Pauli flow of the diagram including the measurement outcomes,
\begin{align}
    \raisebox{-0.4\height}{\includegraphics[height=5cm]{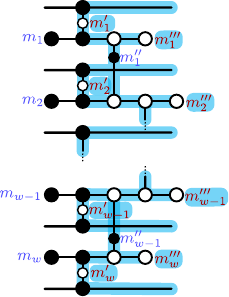}},
\end{align}
and find that that the projector applied is $\Pi_{Z^{\otimes w}}(m)$ with $m=\sum_{i=1}^w (m_i' + m_i''') \mod 2$.

\subsubsection{Depth-6 scheme, halving the space overhead}
The scheme in this section is obtained from the graph
\begin{align}
    \raisebox{-0.4\height}{\includegraphics[height=4cm]{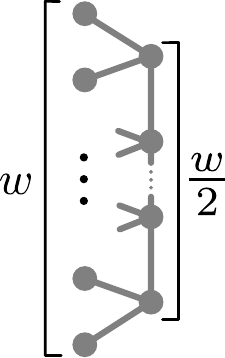}}\qcomma \text{for }w\text{ even and }\quad \raisebox{-0.4\height}{\includegraphics[height=4cm]{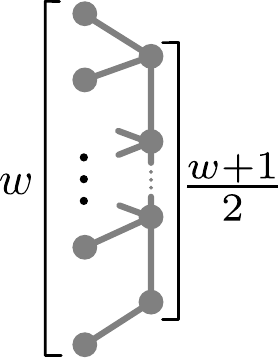}}\qcomma\text{for }w\text{ odd}.
\end{align}
Importantly, each auxiliary node (set of nodes on the right) is connected to not more than two physical nodes (set of nodes on the left).
For now, assume $w$ is even such that every auxiliary node is connected to exactly two physical qubits.
Following the recipe in Sec.\,\ref{sec:connectivity} the graph corresponds to a measurement sequence with the ZX diagram
\begin{align}\label{eq:constant-depth2}
    \raisebox{-0.4\height}{\includegraphics[height=6cm]{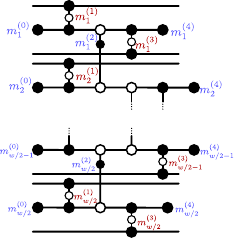}}.
\end{align}
It corresponds to the measurement sequence
\begin{enumerate}
    \item For each auxiliary qubit $a$, measure $X_a$, get outcomes $\{m_i^{(0)}\}_i$.
    \item For each auxiliary qubit $a$, take one the physical qubits it is connected to, $a_1$, and measure $Z_aZ_{a_1}$, get outcomes $\{m_i^{(1)}\}_i$.
    \item Measure $X_aX_{a'}$ between neighboring auxiliary qubits, get outcomes $\{m_i^{(2)}\}_i$.
    \item For each auxiliary qubit $a$, take the other physical qubit it is connected to, $a_2$, and measure $Z_aZ_{a_2}$, get outcomes $\{m_i^{(3)}\}_i$.
    \item For each auxiliary qubit $a$, measure $X_a$.
\end{enumerate}
Again, using Pauli flow or -- equivalently -- a stabilizer tableau simulation we find that the sequence implements the projector $\Pi_{Z^{\otimes w}}(m)$ with $m=\sum_i (m_i^{(1)}+ m_i^{(3)})$.
For $w$ odd the only difference is that the last auxiliary qubit, only connected to a single phyiscal qubit get measured in the $Z$ individually instead of participating in a second $ZZ$ measurement in step 4.

\subsection{Interpolating schemes}\label{sec:interpolating}
In this section we present a scheme interpolating between the two schemes presented in the previous sections.
Specifically, we give a DWR$_2$ scheme using $a$ auxiliary qubits for any integer $2\leq a\leq w$ with a spacetime overhead of $\Theta(w)$.
The two examples above can be understood as special cases of the construction for $a=2,\lceil w/2\rceil, w$.

The DWR$_2$ using $a$ auxiliary qubits is derived from the following graph:
\begin{align}\label{eq:interpolating_graph}
    \raisebox{-0.4\height}{\includegraphics[height=4cm]{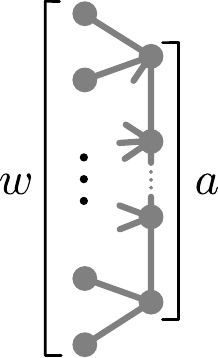}}.
\end{align}
Note that we only sketch the edges touching the auxiliary qubits. In general, the degree of the auxiliary qubits does not have to be the same across all auxiliary qubits.
More precisely, the graph that defines the DWR with $a$ auxiliary qubits is built in the following way:
\begin{enumerate}
    \item Take $w+a$ nodes, labelled from $1$ to $w+a$. The first $w$ nodes will play the role of leaves of the graph (aligned to the left in the above figure) and the remaining $a$ the role of their parents (aligned to the right of the above figure).
    \item Connect the auxiliary nodes along a line: for $j=2,3,...,a-1$, auxiliary node  $w+j$ is connected to $w+j\pm 1$. Node $w+1$ is connected to node $w+2$ and $w+a$ to $w+a-1$, respectively.
    \item Connect node $i=1,2,..,w$ with node $w+1+(i\mod a)$.
\end{enumerate}
This graph results in a DWR scheme with $a$ auxiliary qubits and a depth that behaves quite erratically in both $a$ and $w$.
We obtain a constant depth DWR for $a\geq w/2$ with the minimum depth of $5$ for $a=w$ and depth $6$ for the other cases, cf. Sec.\,\ref{sec:constant_time}.
For smaller, but fixed, $a$, we find that increasing $w$ by 1 either does not change the depth or increases it by 1 or 4.\footnote{Note that the scheme in Sec.\,\ref{sec:constant_space} has a similar behavior, where the depth only increases in multiples of 4, a special case of the $a=2$ case.}
This increase happens when $w$ is increased in multiples of $a-1$ or $a$, respectively, leading to a linear scaling of the depth with $w/a$.
In App.\,\ref{app:depth} we explain some of the behavior of the depth with respect to $a$ and $w$.
We hence obtain a full spacetime overhead linear in $w$, independent of the number of auxiliary qubits.

\subsection{Incompressible spacetime volume}
In the schemes presented above we can calculate a \textit{spacetime volume} $V = AD$ needed to perform the measurement, where $A$ is the number of auxiliary qubits needed and $D$ the depth of the physical measurement circuit.
We find that for the examples above there is a spacetime overhead of $\Theta(w)$.
In the following, we want to argue that this scaling is unavoidable in any compilation in terms of low-weight operations.

Any implementation of $\Pi_P$ in terms of a circuit where each element only involves a maximum of two qubits can be obtained by applying ZX rewrite rules to the tensors in an abstract diagram of the form
\begin{align}
    \raisebox{-0.4\height}{\includegraphics[height=2cm]{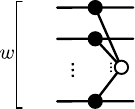}},
\end{align}
representing the projector associated to the measurement.
In this diagram there are $w$ contracted edges (connecting $\delta$- and $\oplus$-tensor).
After rewriting, each edge corresponds to a 2-qubit entangling gate between the auxiliary system used to implement the measurement and the physical qubits.
This includes both compilations in terms of unitary entangling gates, such as $CX$ gates, or in terms of measurements, such as the DWR schemes constructed in this paper.
Assuming each qubit in the auxiliary system can only participate in a single entangling operation at the same time, $A$ bounds the number of these operations per step.
In order to apply all $w$ entangling operations needed we hence need a depth of at least $D\geq w/A$.
Taken together, this bounds the volume to $V \geq A w/A = w$.

In practice we find that unitary implementations using only a single auxiliary qubit are the closest to saturating this bound by using exactly $w$ $CX$ gates and two single-qubit measurements.
The DWR$_2$ schemes in this paper also achieve a linear scaling but with prefactors ranging from $3$ to $5$.
It would be interesting to see if the loose bound above can be tightened for the special case where every entangling operation is a measurement, either by constructing more efficient DWR$_2$s or proving a larger lower bound.

\subsection{Parallelizing DWRs}
Throughout this section we have considered a single Pauli measurement.
In practice, for example in an active quantum error-correcting protocol, many measurements have to be performed after each other, possibly involving the same qubits in the device.
This raises the question to which extent it is possible to compress the spacetime overhead of repeated measurements by ``interleaving'' operations from different measurement gadgets.
This idea is not new and has been explored in the context of specific read-out circuits using unitary entangling gates for specific read-out circuits of error-correcting codes~\cite{Tomita2014Low-distance, Fowler2012Surface, Conrad2018, Geher2024}.
In this section we want to comment on the tradeoffs in parallelizing the schemes presented above. While most considerations are done with DWRs in mind they also apply to other compilations into weight-2 circuit elements.

Consider two consecutive (weight-$w$) measurements that overlap on a set of qubits $O$.
Na\"ively one may be tempted to say that the spacetime overhead of performing both measurements is 2 times the overhead of an individual measurement.
This is far from true: Once a qubit in $O$ has participated in the two-qubit gate entangling it with the auxiliary qubits of the first measurements it is free to participate in the second measurement. That way, Ref.~\cite{GransSamuelsson2024improvedpairwise}, for example, constructed DWR$_2$ schemes of depth 4 that overlap on 2 qubits but can be performed fully in parallel.

Let us shortly outline the up- and downsides of the two extreme cases presented in Secs.\,\ref{sec:constant_space} and \ref{sec:constant_time} in terms of parallelizability.
The first scheme uses a constant number (2) of auxiliary qubits for the measurement but takes time linear in $w$.
Each physical qubit, however, idles for the majority of the time.
Hence, allowing for 2 auxiliary qubits per measurement, we can in principle perform $O(w)$ measurements that involve the qubits on which the measurement overlaps.
In this scenario, each of the $O(w)$ measurements will be finished at the same time.
In the second scheme, since the depth of the DWR is constant (5 or 6) the idling time of each physical qubits is also constant (4 or 5) and by having $w$ auxiliary qubits per measurement we could parallelize 4, respectively 5, measurements.
This comes at a big cost of auxiliary qubits.
At the same time, the constant depth scheme offers the opportunity to reuse auxiliary qubits after they are measured out.
This includes both the auxiliary qubits that are measured out after 4 or 5 steps in the circuit shown in \,\eqref{eq:constant-depth} and \eqref{eq:constant-depth2} and the ones that are read out one timestep later.
By mirroring the worldlines of the auxiliary qubits in the circuit of \,\eqref{eq:constant-depth} we find that repeated measurements can be compressed.
Similarly, the auxiliary qubits in the circuit \eqref{eq:constant-depth2} can directly be reused.
The measurement used for the read-out can be thought of an initialization for the next block.

To conclude, the schemes we explored in this section serve as two extreme cases of DWRs.
The first scheme takes time linear in the weight of the measurement performed by the DWR but offers a large flexibility for parallelization, namely extensively in the weight of the measurements that we want to perform.
In this scenario, the measurement outcomes of the high-weight measurements are gathered roughly at the same time.
This makes them potentially worse performing in the context of error correction due to the many fault locations within the densely parallelized circuits and might be a guiding strategy for compilation on the logical level, where qubits are more expensive in resources and one might want to implement Pauli measurements of very high weight.
The second scheme performs the weight-$w$ measurement in constant time and hence does not need to be parallelized as much.
In stark contrast to the previous scheme, the rate at which the outcomes of these measurements can be determined is constant.
This makes DWR schemes using more auxiliary qubits more promising for lower-level implementations, for example, for syndrome extraction within a QEC scheme.
In order to achieve a dense circuit, reusing auxiliary qubits is particularly helpful to reduce the total overhead of repeated measurements.

\section{Concluding remarks}
We have given arguments why the scaling of the spacetime overhead to implement a weight-$w$ measurement scales linearly with $w$ if compiled into a sequence of two-qubit operations.
It would be interesting to devise a more fine-grained quantity of overheads needed for specific DWR schemes.
Depending on the physical implementation of the measurements different trade-offs that also include idling times of individual qubits might be useful.

So far, we have explored DWRs where the subset of qubits that play the role of physical qubits is fixed.
Within the ZX calculus it is straightforward to devise schemes where the logical information is teleported between different qubits and each physical, or auxiliary, qubit only needs to be alive for a constant time.
E.g., the DWR scheme constructed in Sec.\,\ref{sec:constant_space} could be adapted such that auxiliary qubit $a$ and $b$ change roles every four timesteps.
It would be interesting to explore how these implementations can be used to devise noise-reduction schemes similar to the one presented in Ref.~\cite{delfosse2024clinr}.

Furthermore, exploring the possibility of interleaving, respectively parallelizing, DWRs should be important when incorporating them into subroutines of useful protocols and algorithms.
In the context of fault-tolerant implementations of algorithms it would be interesting to explore to which extend there are different trade-offs on the physical and on the logical level.
For example, in order to ensure that error syndromes are gathered fast enough to perform reliable decoding, a minimal amount of auxiliary qubits might be needed. On the logical level, on the other hand, one might have more freedom to parallelize DWRs to reduce the total runtime of a routine that involves mid-circuit measurements.

The treatment of hook errors and more general error-equivalences within DWR circuits is more intricate than within unitary implementations of a read-out circuit.
Ref~\cite{GransSamuelsson2024improvedpairwise} constructed a DWR scheme for surface-code stabilizers that preserve the fault distance of the underlying code.
A more general construction of circuits minimizing error-spread beyond individual examples is lacking and important for the implementation of qLDPC codes~\cite{Breuckmann2021qLDPC} that go beyond variants of the surface code~\cite{Fowler2012Surface}.
We believe that deformations of ZX diagrams together with Pauli flows can be useful in the understanding of properties of a read-out circuit that are responsible for the effective fault distance of a read-out circuit.
It would be interesting to incorporate existing approaches to include fault tolerance into a setting with restricted connectivity, such as, e.g., the approach presented in Ref.~\cite{Lao_2020}, into DWRs and to compare the construction with other dynamical weight-reduction schemes for stabilizer read-out, cf. Ref.~\cite{baspin2024wirecodes}.

\emph{Note added:}
During the preparation of this manuscript I was made aware of two related works: \cite{baspin2024wirecodes} provides DWR$_3$ schemes for the read-out of the generators of a stabilizer group and \cite{distance-rewrite} constructs distance-preserving rewrite rules for circuits representing broad families of QEC codes.

\section*{Acknowledgements}
I want to thank D. Williamson for inspiring discussions that initiated the constructions presented in this paper and both the authors of Ref.\,\cite{baspin2024wirecodes} and \cite{distance-rewrite} for informing me about their work.
I am grateful to A. Quintavalle for helpful feedback on earlier stages of the draft and clarifying discussions that greatly improved the paper. 
I want to thank A. Townsend-Teague, J. Conrad, J. Old and J. Eisert for constructive feedback on the draft.
Moreover, I am thankful to J. Eisert for his continuous support of independent research and for creating an atmosphere in which this research was possible.
I acknowledge support from the DFG (CRC 183).

\bibliographystyle{unsrt}
\bibliography{measurements}

\newpage 

\begin{appendix}
\section{DWRs from deforming ZX diagrams}\label{app:constant-depth-splitting}
In this section we present how the DWRs presented in Secs.\,\ref{sec:constant_space} and \ref{sec:constant_time} can be obtained from deformations of ZX diagrams, as defined in Sec.\,\ref{sec:method}.

Starting from the post-selected $m=0$ diagram from \eqref{eq:P-measurement-blackboxZX}, we perform the following splitting operations:
\begin{subequations}\label{eq:splitting-constant-space}
    \begin{align}
        \raisebox{-0.4\height}{\includegraphics[height=3.5cm]{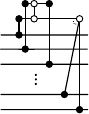}} \;=& \; \raisebox{-0.4\height}{\includegraphics[height=3.5cm]{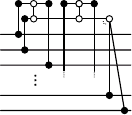}} \;=\; \raisebox{-0.4\height}{\includegraphics[height=3.5cm]{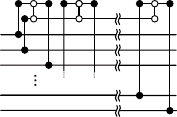}}\\
        =& \; \raisebox{-0.4\height}{\includegraphics[height=3.5cm]{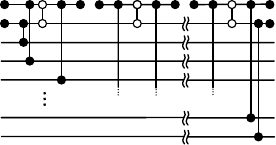}} \;.
    \end{align}        
\end{subequations}
To obtain a post-selected version of the measurement scheme presented in Sec.\,\ref{sec:constant_space}.

By choosing a different deformation we can implement the same measurement with a constant depth.
Specifically, we can split tensors in \eqref{eq:P-measurement-blackboxZX} to get
\begin{align}
    \raisebox{-0.4\height}{\includegraphics[height=5cm]{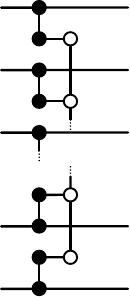}} = \raisebox{-0.4\height}{\includegraphics[height=5cm]{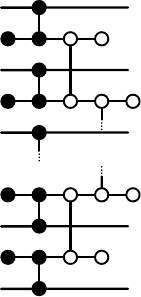}} .
\end{align}

\section{Depth of DWRs from Sec.\,\ref{sec:interpolating}}\label{app:depth}
In this appendix we want to explain how the depth of the family of DWRs constructed in Sec.\,\ref{sec:interpolating} depends on $a$ and $w$.
First of all, we note that $D$ is constant for $a\geq \lceil w/2\rceil$, cf. Sec.\,\ref{sec:constant_time}.

Hence, the depth increases with $w$ only when $a<w/2$.
In the following, we consider a DWR obtained for fixed $w$ and $a$ and describe how the depth changes when changing $w$ while keeping $a$ fixed.
We start with the case where $w=2a+m(a-1)$ for some $m\in\bZ^+$.
In that case the DWR ends with a round of $ZZ$ and $X$ measurements on every qubit.
The associated ZX diagram for the last two rounds (on the auxiliary qubits) looks like
\begin{align}\label{app:eq:starting_DWR_1}
    \raisebox{-0.4\height}{\includegraphics[height=3.5cm]{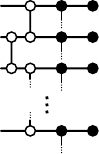}}.
\end{align}
The connections between $\oplus$-tensors are all among auxiliary qubits and the connections emanating vertically from the $\delta$-tensors correspond to $ZZ$ measurements with physical qubits which are not shown in the diagram.
Next, we consider a DWR constructed for the same $a$ but we increase $w\mapsto w+1$.
This will affect the DWR such that the end now looks like
\begin{align}
    \raisebox{-0.4\height}{\includegraphics[height=3cm]{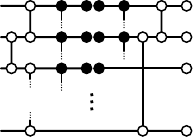}}
\end{align}
We find that in this case the depth was increased by $4$, assuming $a>2$, where the $XX$ measurements at the second-to-last step take a depth of 2 due to their overlap.
Increasing $w$ further does not change the depth until it was increased by $a-1$ compared to the weight that lead to the diagram \eqref{app:eq:starting_DWR_1}.
At this stage, the DWR ends with
\begin{align}
    \raisebox{-0.4\height}{\includegraphics[height=3.5cm]{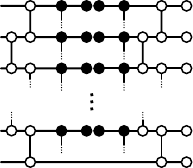}}.
\end{align}
and increasing $w$ further by 1 increases the depth by only 1, leading to a DWR that ends with
\begin{align}
    \raisebox{-0.4\height}{\includegraphics[height=3.5cm]{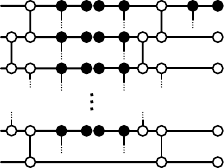}}.
\end{align}
From here, increasing $w$ further by $a-1$ does not change the depth and we obtain a DWR that again ends with a measurement sequence like in \eqref{app:eq:starting_DWR_1}.



\end{appendix}
\end{document}